\documentclass[reprint,twocolumn,showpacs,amsmath,amssymb,aps,pra,preprintnumbers]{revtex4-1}

\usepackage{graphicx}
\usepackage{dcolumn}
\usepackage{bm}
\usepackage[T1]{fontenc}
\newcommand{\rozmiar}{0.47\textwidth}
\newcommand{\rozmiardwa}{0.47\textwidth}
\newcommand{\rozmiartrzy}{0.47\textwidth}
\begin{document}

\preprint{Submitted to: ACTA PHYSICA POLONICA A}

\title{Some properties of the model of a~superconductor\\ with pair hopping and magnetic interactions at half-filling}

\author{Konrad Kapcia}%
    \email[e-mail: ]{kakonrad@amu.edu.pl}
\affiliation{%
Electron States of Solids Division,
Faculty of Physics, Adam Mickiewicz University, Umultowska 85, 61-614 Pozna\'n, Poland}

\date{November 25, 2011}


\begin{abstract}
We present our preliminary studies of an~effective model of a~superconductor with short coherence length involving magnetic interactions. The Hamiltonian considered consists of (i)~the effective on-site interaction~$U$, (ii)~the intersite magnetic exchange interactions ($J^z$, $J^{xy}$) between  nearest-neighbors and (iii)~the intersite charge exchange term~$I$, determining the hopping of electron pairs between nearest-neighbor sites.
In the analysis of the phase diagrams and thermodynamic properties of this model for half-filling (\mbox{$n=1$}) we have adopted the variational approach, which treats the on-site interaction term exactly and the intersite interactions within the mean-field approximation.
One finds that the system considered can exhibit very interesting multicritical behaviors (including tricritical, critical-end and bicritical points) caused by the competition between magnetism and superconductivity, even for \mbox{$n=1$}.
Our investigations show that, depending on the values of interaction parameters, the system  at half-filling can exhibit three homogeneous phases: superconducting (SS), (anti-)ferromagnetic (F) and nonordered (NO). The transitions between ordered phases (SS, F) and the NO phase can be first order as well as second order ones, whereas \mbox{SS--F} transition is first order one.
Temperature dependencies of the order parameters and thermodynamic properties of the system at the sequence of transitions: SS$\rightarrow$F$\rightarrow$NO with increasing temperature for \mbox{$J/I=0.3$}, \mbox{$U/I_0 = 0.69$} and \mbox{$n=1$} are also presented.
\end{abstract}




\maketitle


\section{Introduction}

There has been much interest in superconductivity with very short coherence length. This interest is due to its possible relevance to high temperature superconductors (the cuprates, doped bismuthates, fullerenes and iron-based) and also to the several other exotic superconducting materials  (for a review, see Refs.~\onlinecite{MRR1990,AAS2010} and references therein). It can also give relevant insight into a~behavior of strongly bounded fermion pairs on the optical lattices.

The interplay and competition between superconductivity and magnetic orderings is currently under intense investigations (among others in iron chalcogenides and cuprates, e.~g. Refs.~\onlinecite{PIN2009,RPC2011,XTP2008} and references therein). A~conceptually simple model for studying that competition will be studied in this report.

The Hamiltonian considered has the following form:
\begin{eqnarray}\label{row:ham1}
\hat{H} & = & U\sum_{i}{\hat{n}_{i\uparrow}\hat{n}_{i\downarrow}}- I\sum_{\langle i,j\rangle}{\left(\hat{\rho}_i^+\hat{\rho}_j^- + \hat{\rho}_j^+\hat{\rho}_i^-\right)} + \\
& - & 2J\sum_{\langle i,j \rangle }{ \hat{s}_i^z\hat{s}^z_j} - \mu\sum_i\hat{n}_i, \nonumber
\end{eqnarray}
where  \mbox{$\hat{n}_{i}=\sum_{\sigma}{\hat{n}_{i\sigma}}$}, \mbox{$\hat{n}_{i\sigma}=\hat{c}^{+}_{i\sigma}\hat{c}_{i\sigma}$}, \mbox{$\hat{\rho}^+_i=(\hat{\rho}^-_i)^\dag=\hat{c}^+_{i\uparrow}\hat{c}^+_{i\downarrow}$},
\mbox{$\hat{s}_i^z=(1/2)(\hat{n}_{i\uparrow} - \hat{n}_{i\downarrow})$}. $\hat{c}_{i\sigma}$ and $\hat{c}^{+}_{i\sigma}$ denote annihilation and creation operators of an electron with spin \mbox{$\sigma=\uparrow,\downarrow$} at the site $i$,
which satisfy canonical anticommutation relations
\begin{equation}\label{row:antycomutation}
\{ \hat{c}_{i\sigma}, \hat{c}^+_{j\sigma'}\} = \delta_{ij}\delta_{\sigma\sigma'}, \quad
\{ \hat{c}_{i\sigma}, \hat{c}_{j\sigma'}\} = \{ \hat{c}^+_{i\sigma}, \hat{c}^+_{j\sigma'}\} = 0,
\end{equation}
where $\delta_{ij}$ is the Kronecker delta.
\mbox{$\sum_{\langle i,j\rangle}$} indicates the sum over nearest-neighbor sites $i$ and $j$ independently.
$U$ is the on-site density interaction,
$I$ is the intersite  charge  exchange interaction between nearest neighbors and
$J$ is the Ising-like magnetic interaction between nearest neighbors.
$\mu$ is the chemical potential, depending on the concentration
of electrons:
\begin{equation}\label{row:condn1}
n = \frac{1}{N}\sum_{i}{\left\langle \hat{n}_{i} \right\rangle},
\end{equation}
with \mbox{$0\leq n \leq 2$} and $N$ is the total number of lattice sites.

There are two competitive interaction parameters of the model: (i)~the pair hopping interaction $I$, determining the electron pair mobility and responsible  for the long-range superconducting order (local pairing mechanism) and
(ii)~the Ising-like interaction $J$ between nearest neighbors responsible for magnetism in the system.
The on-site density-density interaction $U$ contributes (together with $I$) to the pair binding energy by reducing (\mbox{$U>0$}) or enhancing (\mbox{$U<0$}) its value. Moreover, repulsive \mbox{$U>0$} favors magnetic ordering.
To simplify our analysis we do not include in Hamiltonian (\ref{row:ham1}) the single electron hopping term ($\sum_{i,j}t_{ij}\hat{c}^{+}_{i\sigma}\hat{c}_{j\sigma}$) as well as other inter-site interaction terms. This assumption corresponds to the situation when single particle mobility is much  smaller than the pair mobility and can be neglected.

The interactions $U$, $I$ and $J$ will be treated as the effective ones and will be assumed to include all the possible contributions and renormalizations like those coming from the strong electron-phonon coupling or from the coupling between electrons and other electronic subsystems in solid or chemical complexes \cite{MRR1990}.

Ferromagnetic XY-order of pseudospins $\hat{\vec{\rho}}_i$ (for \mbox{$I>0$}) corresponds to the SS phase ($s$-pairing superconducting), whereas the antiferromagnetic XY-order (for \mbox{$I<0$}) -- to the S$\eta$ phase ($\eta$-pairing superconducting). For \mbox{$t_{ij}=0$} there is a well known isomorphism between the SS and S$\eta$ cases (with an obvious redefinition of the order parameter: \mbox{$\Delta=\Delta_{SS} = \frac{1}{N}\sum_i{\langle \hat{\rho}^-_i\rangle}$}, for \mbox{$I>0$} and \mbox{$\Delta_{\eta S} = \frac{1}{N}\sum_i{\exp{(\textrm{\textbf{i}}\vec{Q}\cdot\vec{R}_i)}\langle \hat{\rho}^-_i\rangle} $}, for \mbox{$I<0$}, $\vec{Q}$ is half of the smallest reciprocal lattice vector)
for lattices consisting of two interpenetrating sublattices such as for example SC or BCC lattices.
One should also notice that, in the absence of the single electron hopping term, ferromagnetic (\mbox{$J > 0$}) interactions are simply mapped onto the antiferromagnetic cases (\mbox{$J < 0$}) by redefining the spin direction on one sublattice in lattices decomposed into two interpenetrating sublattices. Thus, we restrict ourselves to the case of \mbox{$I>0$} and \mbox{$J>0$}.

We have performed extensive study of the phase diagrams of model (\ref{row:ham1}) for arbitrary $n$ and $\mu$~\cite{KR0000}. In the analysis we have adopted a variational approach (VA), which treats the onsite interaction term ($U$) exactly and the intersite interactions ($I$, $J$) within the mean-field approximation (MFA).
In this paper we present our preliminary results for the half-filling (\mbox{$n=1$}).

Model (\ref{row:ham1}) has been analyzed within VA only for particular cases: (i)~\mbox{$J=0$}~\cite{RP1993,R1994,B1973,HB1977,KRM0000} and (ii)~\mbox{$I=0$}~\cite{KKR2010,MKPR2012} till now. The rigorous results for \mbox{$I=0$} in ground state have been also obtained~\cite{BS1986}.
Some preliminary study of the \mbox{$I=0$} case in finite temperatures using Monte Carlo simulations has also been done for a~square lattice~\cite{MKPR2012}.

The ferromagnetic (F) phase is characterized by nonzero value of the magnetic order parameter (magnetization) defined as \mbox{$m =(1/N)\sum_i{\langle\hat{s}^z_i\rangle}$} (and \mbox{$\Delta=0$}), in the superconducting (SS) phase the order parameter \mbox{$\Delta \neq 0$} (and \mbox{$m=0$}) and in the nonordered (NO) phase \mbox{$m = 0$} and \mbox{$\Delta=0$}.

Within the VA the intersite interactions are decoupled within the MFA, what let us find a~grand canonical potential per site $\omega(\mu)$ (or free energy per site \mbox{$f(n)=\omega(\mu)+\mu n$}) in the grand canonical ensemble. One can also calculate the averages: $n$, $\Delta$ and $m$, what gives a~set of three non-linear self-consistent equations (for homogeneous phases).
This set for \mbox{$T>0$} is solved numerically and one obtains $\Delta$, $m$, and $n$ (or $\mu$) when $\mu$ (or $n$) is fixed. It is important to find a solution corresponding to the lowest $\omega(\mu)$ (or $f(n)$).
For \mbox{$n=1$} one obtains \mbox{$\mu = U/2$} and two equations for $\Delta$ and $m$ need to be solved numerically.

We also introduce the following denotation: \mbox{$I_0 = z I$}, \mbox{$J_0 = z J$},
where $z$ is the number of nearest neighbors.

\section{Results and discussion}

\begin{figure}
    \centering
    \includegraphics[width=\rozmiar]{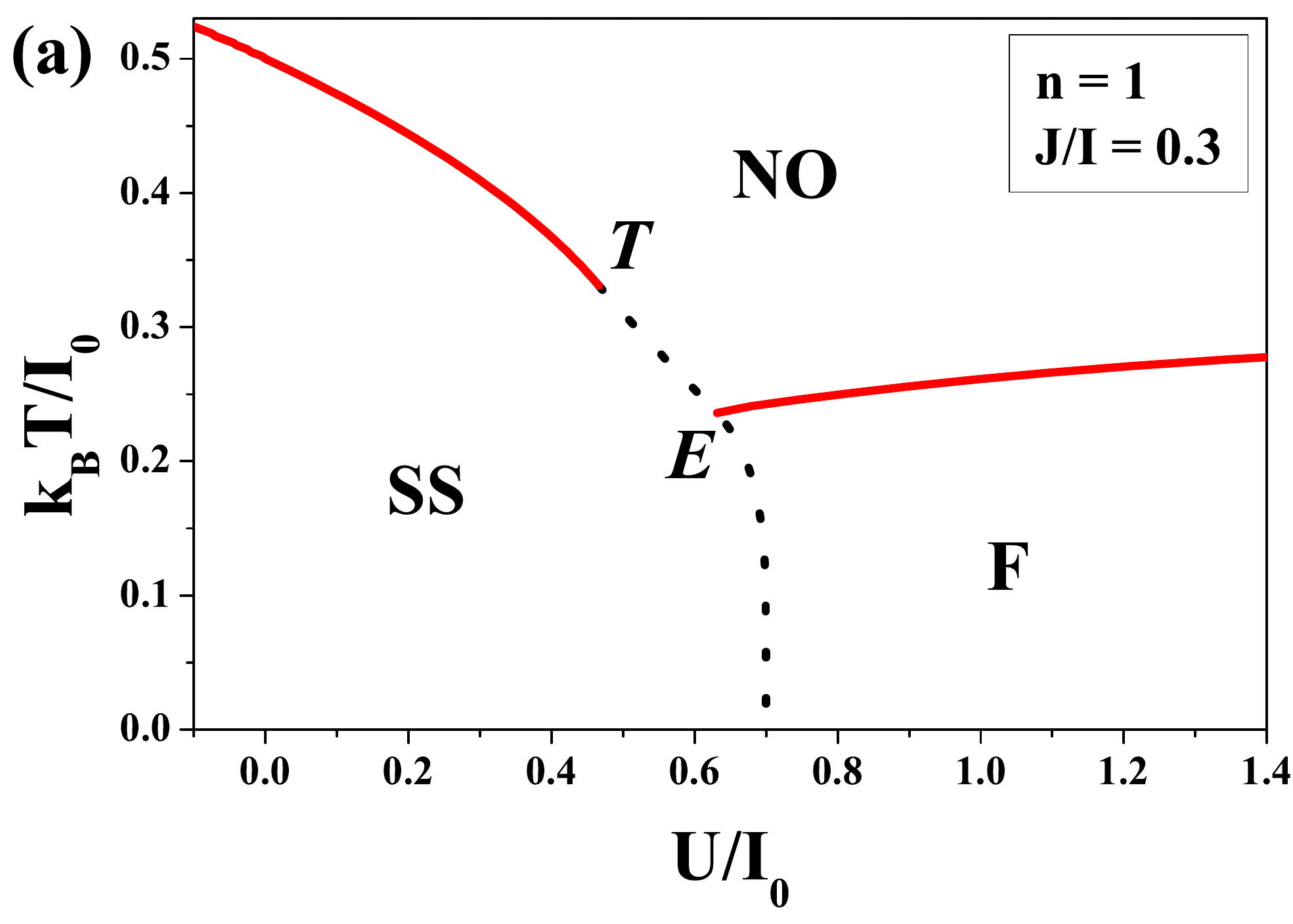}
    \includegraphics[width=\rozmiar]{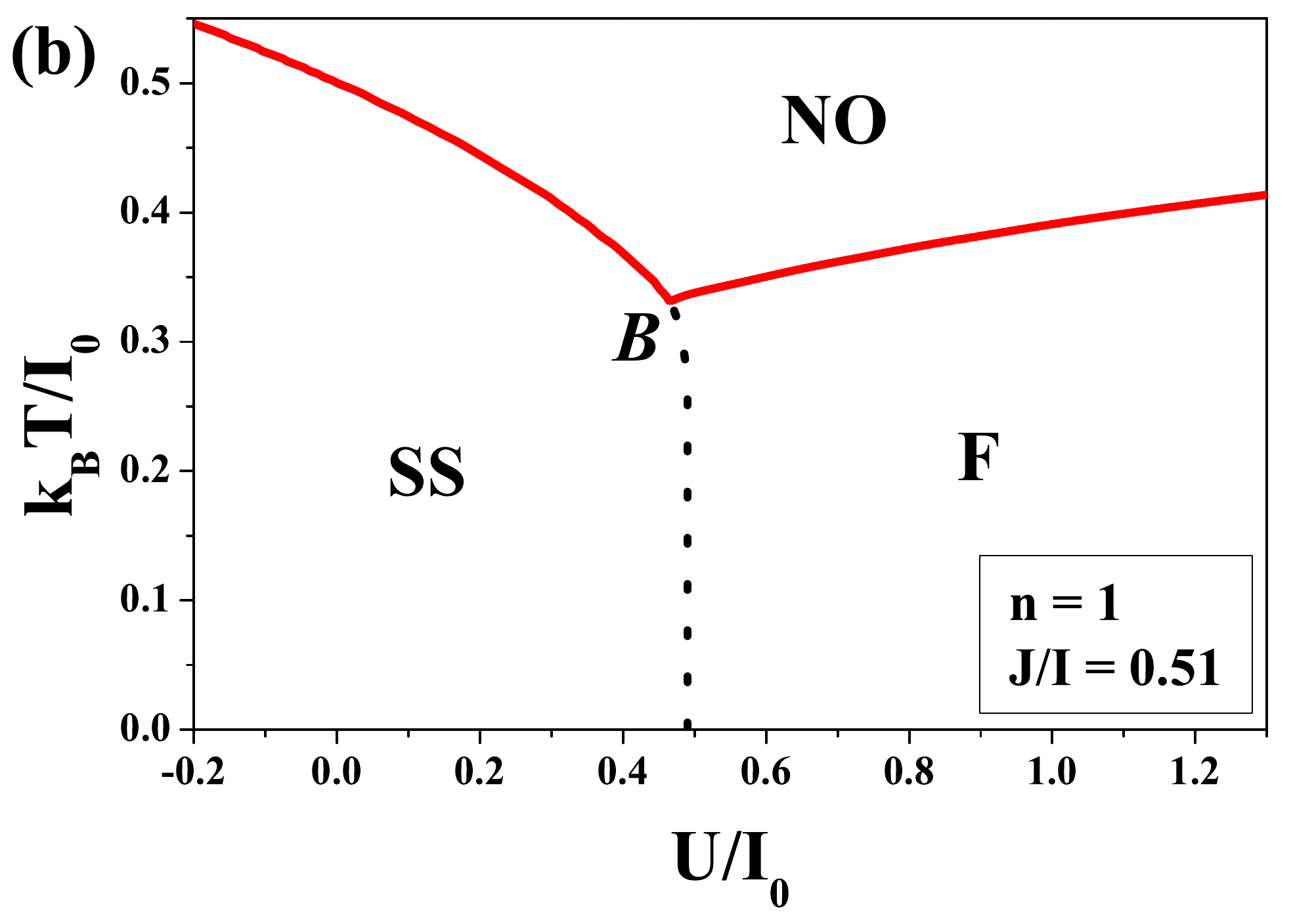}
    \includegraphics[width=\rozmiar]{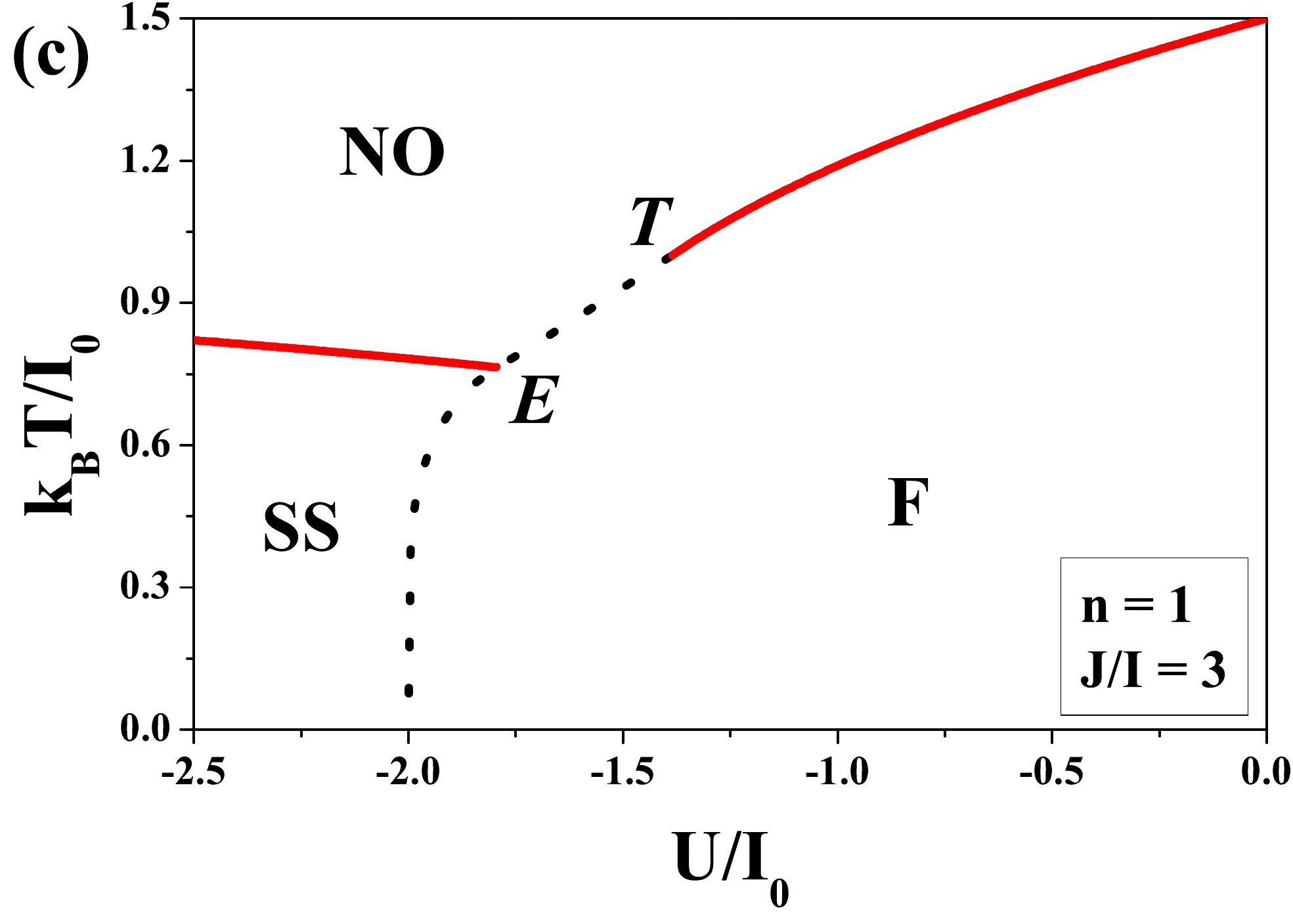}
    \caption{(Color online) Phase diagrams $k_BT/I_0$~vs.~$U/I_0$ at half-filling (\mbox{$n=1$}) for (a)~\mbox{$J/I=0.3$}, (b)~\mbox{$J/I=0.51$} and (c)~\mbox{$J/I=3$}. Dotted and solid lines indicate first order and second order boundaries, respectively. $T$, $E$ and $B$ denote tricritical, critical-end and bicritical points, respectively.}
    \label{rys:diagramy}
\end{figure}

There are two well defined limits of model~(\ref{row:ham1}): (i)~\mbox{$U\rightarrow-\infty$} favoring superconductivity 
and (ii)~\mbox{$U\rightarrow+\infty$}, where only magnetic orderings can appear in the system.

For \mbox{$U\rightarrow-\infty$} (states with single occupancy  are excluded and only local pairs can exists in the system) the model is equivalent with the hard-core charged boson model on the lattice~\cite{RP1993,MRK1995,BBM2002}. In this limit the \mbox{SS--NO} transition
is second order one and is to the NO phase being a~state of dynamically disordered local pairs. The \mbox{SS--NO} transition temperatures increase monotonically with decreasing \mbox{$|n-1|$}. The maximum value of the transition temperature is \mbox{$k_BT/I_0=1$} for \mbox{$n=1$}~\cite{RP1993,KRM0000}.

In the opposite limit (i.~e. \mbox{$U\rightarrow+\infty$}) the double occupied sites are excluded and only the magnetic states can occur on the phase diagram~\cite{KKR2010,MKPR2012}. At sufficiently low temperatures the homogeneous phases are not states with the lowest free energy and the PS state are stable (if \mbox{$n\neq1$}). On the phase diagram there is a~second order line at high temperatures, separating the F and NO phases, whereas first order transition takes place at lower temperatures, leading to a phase separation of the F and NO phases. The critical point for the phase separation (tricritical point) lies on the second order \mbox{F--NO} line  and it is located at \mbox{$k_BT/J_0=1/3$} and \mbox{$n=1/3$}~\cite{KKR2010}.
The  \mbox{F--NO} (second order) transition temperature decreases with increasing \mbox{$|n-1|$} and  its maximum value is \mbox{$k_BT/J_0=1$} for \mbox{$n=1$}~\cite{MKPR2012}.

\subsection{The phase diagrams at half-filling}

A few representative $k_BT/I_0$ vs. $U/I_0$ phase diagrams of model~(\ref{row:ham1}) evaluated for various ratios of $J/I$ at half-filling (\mbox{$n=1$}) are presented in Fig.~\ref{rys:diagramy}.

The phase diagram $k_BT/I_0$~vs.~$U/I_0$ for \mbox{$J/I=0.3$} and \mbox{$n=1$} is shown in Fig.~\ref{rys:diagramy}a. Two ordered phases: the SS phase and the F phase are separated by first order boundary on the diagram. Both order parameters change discontinuous at the \mbox{SS--F} transition. With increasing $U/I_0$ the \mbox{SS--NO} transition temperature decreases from \mbox{$k_BT/I_0=1$} at \mbox{$U/I_0\rightarrow-\infty$}. At \mbox{$U/I_0 = \frac{2}{3}\ln (2)\simeq0.462$} and $k_BT/I_0 = 1/3$ the transition changes its type from second order one to first order one resulting in the tricritical point $T$ on the phase diagram. The \mbox{F--NO} transition temperature is slightly dependent on $U/I_0$ and increases to \mbox{$k_BT/I_0 = 0.3$} (\mbox{$k_BT/J_0=1$}) at \mbox{$U/I_0\rightarrow+\infty$}.
The \mbox{F--NO} second order line ends at critical-end point $E$ on the first order boundary of the SS phase occurrence.

The possible sequences of transitions with increasing temperatures and the transition orders of them are listed below (for \mbox{$J/I=0.3$}):
\begin{itemize}
\item[(i)]
SS$\rightarrow$NO: second order, for \mbox{$U/I_0<0.46$} and first order, for \mbox{$0.46<U/I_0<0.63$},
\item[(ii)]
SS$\rightarrow$F$\rightarrow$NO: first order and second order, respectively, for \mbox{$0.63<U/I_0<0.7$},
\item[(iii)]
F$\rightarrow$NO: second order, for \mbox{$U/I_0>0.7$}.
\end{itemize}

The phase diagram for \mbox{$J/I = 0.51$} is qualitatively different than that for \mbox{$J/I = 0.3$}.
For \mbox{$J/I=0.51$} the system exhibits bicritical behavior (Fig.~\ref{rys:diagramy}b) in contrary to the tricritical behavior (and occurrence of $E$-point) for  \mbox{$J/I=0.3$}. Similarly as for \mbox{$J/I=0.3$}, the \mbox{SS--F} transition is first order one while the \mbox{SS--NO} and \mbox{F--NO} transitions are second order ones. The two second order boundaries and the first order boundary merge at bicritical point $B$.

\begin{figure}
    \centering
    \includegraphics[width=\rozmiardwa]{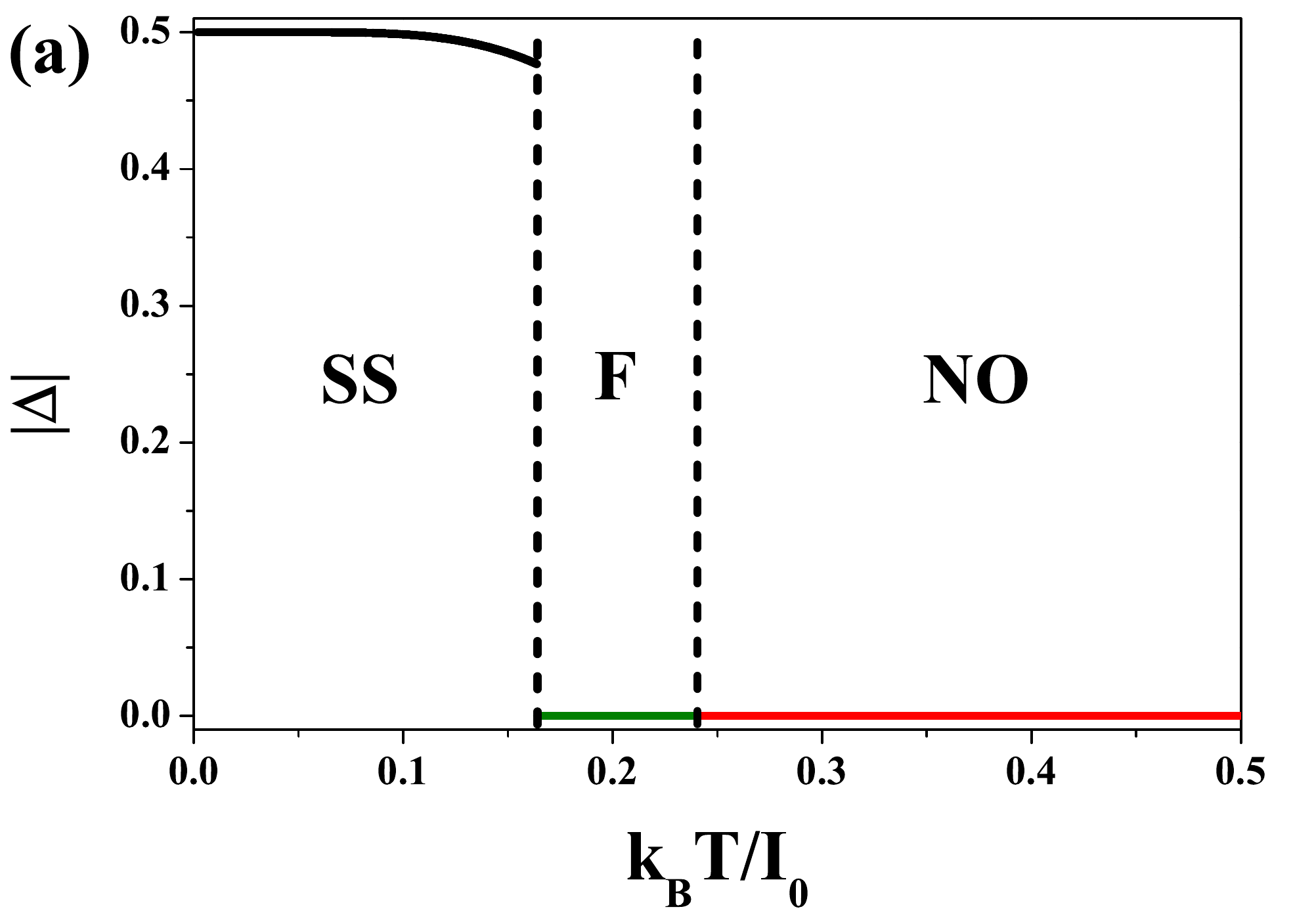}
    \includegraphics[width=\rozmiardwa]{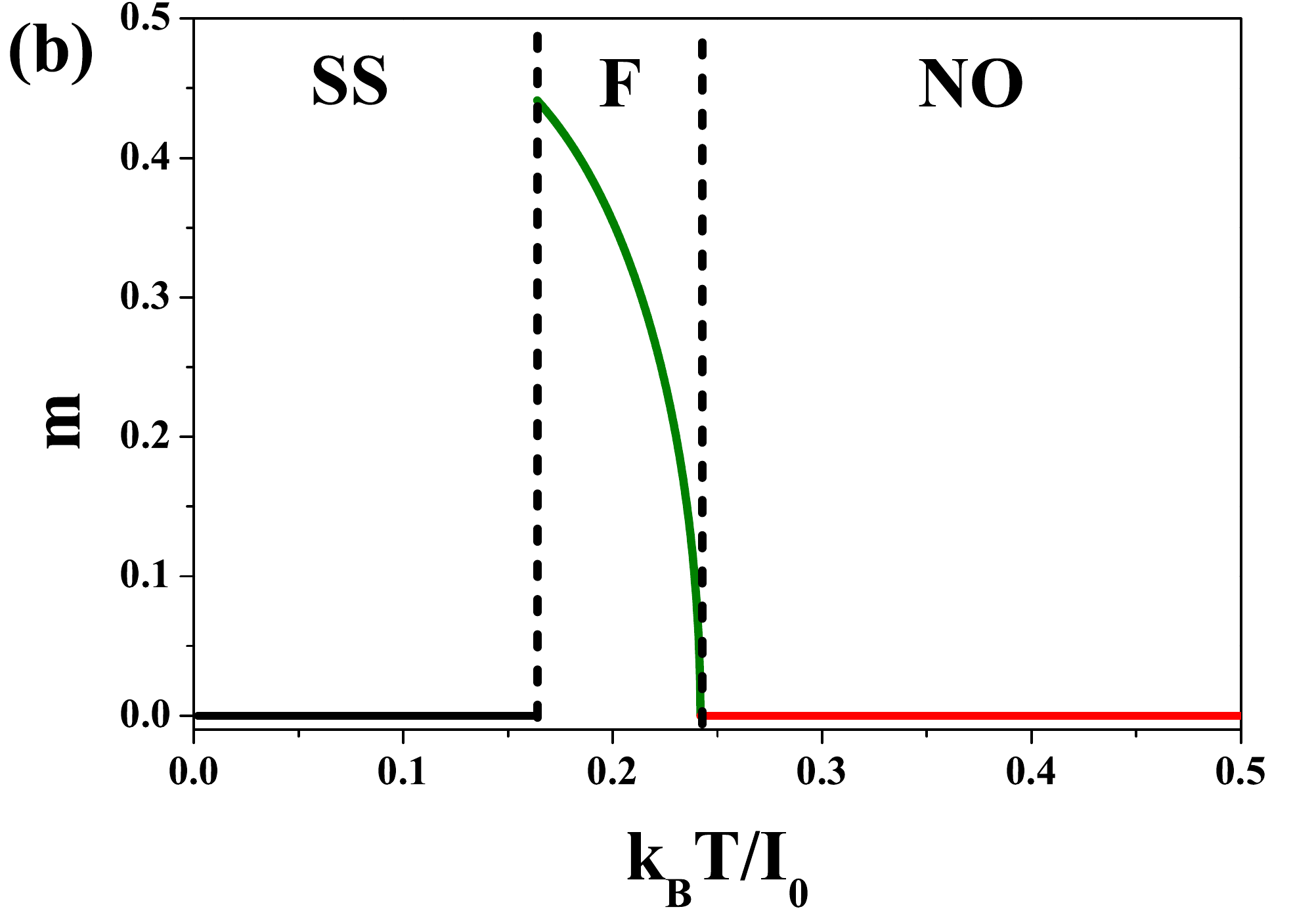}
    \caption{(Color online) Temperature dependence of (a)~superconducting order parameter $|\Delta|$ and (b)~magnetic order parameter $m$  for \mbox{$J/I=0.3$}, \mbox{$U/I_0 = 0.69$} and \mbox{$n=1$}.}
    \label{rys:porzadek}
\end{figure}

The system exhibits the tricritical behavior for \mbox{$J/I<0.5$}, whereas the bicritical behavior occurs for \mbox{$0.5<J/I<2$}. For \mbox{$J/I>2$} the system exhibits tricritical behavior again, however the tricritical point $T$ is located at the \mbox{F--NO} line at \mbox{$U/J_0=-\frac{2}{3}\ln(2)\simeq-0.462$} and \mbox{$k_BT/J_0 = 1/3$} (cf. Fig.~\ref{rys:diagramy}c). For \mbox{$J/I>2$} the \mbox{F--NO} transition  can be second order (for \mbox{$U/J_0>-0.46$}) as well as first order (for \mbox{$U/J_0<-0.46$}). Notice that the axis in Fig.~\ref{rys:diagramy} are normalized by $I_0$, not by $J_0$.

One should notice that, for any $J/I$, with increasing $U/I_0$ the \mbox{SS--NO} transition temperature decreases monotonically from \mbox{$k_BT/I_0=1$} at \mbox{$U\rightarrow-\infty$}, whereas the \mbox{F--NO} transition temperature is an increasing function of $U/I_0$ (to its maximum \mbox{$k_BT/J_0 = 1$} at \mbox{$U\rightarrow+\infty$}).

Let us concentrate now on temperature dependencies of the order parameters and thermodynamic properties of the system at the sequence of transitions: SS$\rightarrow$F$\rightarrow$NO for \mbox{$J/I=0.3$}, \mbox{$U/I_0 = 0.69$} and \mbox{$n=1$}.

\subsection{The order parameters}

The temperature dependencies of the order parameters: $\Delta$ and $m$ for \mbox{$J/I=0.3$}, \mbox{$U/I_0 = 0.69$} and \mbox{$n=1$} are presented in Fig.~\ref{rys:porzadek}. It is clearly seen that at the \mbox{SS--F} transition (at \mbox{$k_{B}T_{c1}/I_0 = 0.16$}) the both  order parameters: superconducting order parameter $\Delta$ and magnetization $m$ change discontinuously. In the SS phase \mbox{$\Delta\neq0$} and \mbox{$m=0$} whereas in the F phase \mbox{$m\neq0$} and \mbox{$\Delta=0$}. The \mbox{F--NO} transition (at \mbox{$k_{B}T_{c2}/I_0 = 0.24$}) is connected with a~continuous decay of $m$ at the transition temperature.

\subsection{The thermodynamic properties}

Calculating the free energy  per site $f$ one can obtain thermodynamic characteristics of the system for arbitrary temperature.
The double occupancy  per site  $D$ is defined as:
\begin{equation}
D = \frac{1}{N}\sum_i \langle \hat{n}_{i\uparrow} \hat{n}_{i\downarrow} \rangle = \left(\frac{\partial f}{\partial U}\right)_{T}
\end{equation}
and it is related with the local magnetic moment $\gamma$ by the following formula:
\begin{eqnarray}
\gamma & = &\frac{1}{N}\sum_i \langle \hat{s}^z_i \rangle =  \frac{1}{2N}\sum_i{\langle |\hat{n}_{i\uparrow}-\hat{n}_{i\downarrow}| \rangle}   \\
& = & \frac{1}{2}n - \sum_i{\langle \hat{n}_{i\uparrow}\hat{n}_{i\downarrow} \rangle} =  \frac{1}{2}n - \left(\frac{\partial f }{\partial U}\right)_{T} = \frac{1}{2}n - D, \nonumber
\end{eqnarray}
because \mbox{$|\hat{n}_{i\uparrow}-\hat{n}_{i\downarrow}|=(\hat{n}_{i\uparrow}-\hat{n}_{i\downarrow})^2 =\hat{n}_{i\uparrow}+\hat{n}_{i\downarrow} - 2\hat{n}_{i\uparrow}\hat{n}_{i\downarrow}$}, \mbox{$\hat{n}^2_{i\sigma}=\hat{n}_{i\sigma}=0,1$}
and $|\hat{n}_{i\uparrow}-\hat{n}_{i\downarrow}|=0,1$.

\begin{figure}
    \centering
    \includegraphics[width=\rozmiartrzy]{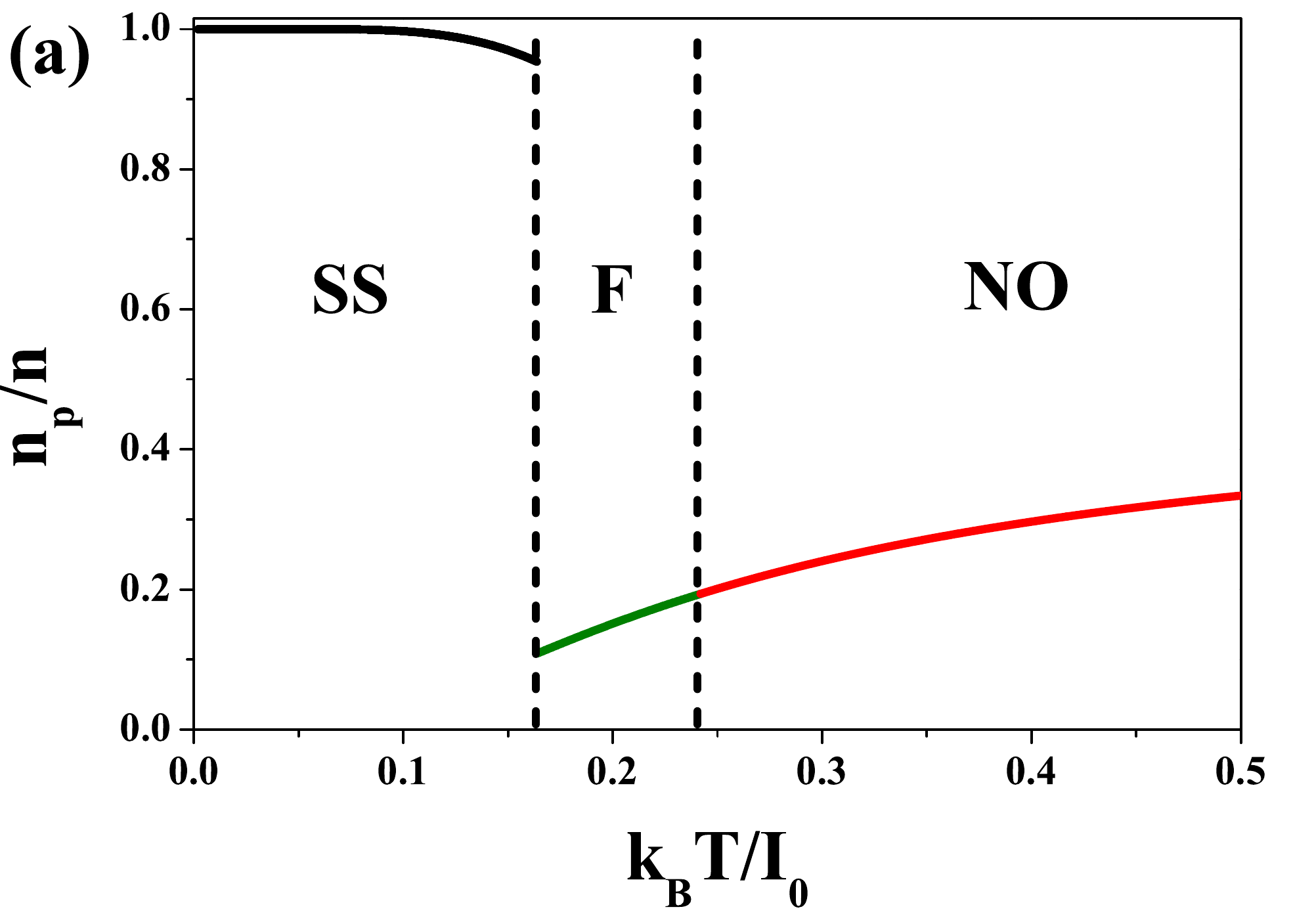}
    \includegraphics[width=\rozmiartrzy]{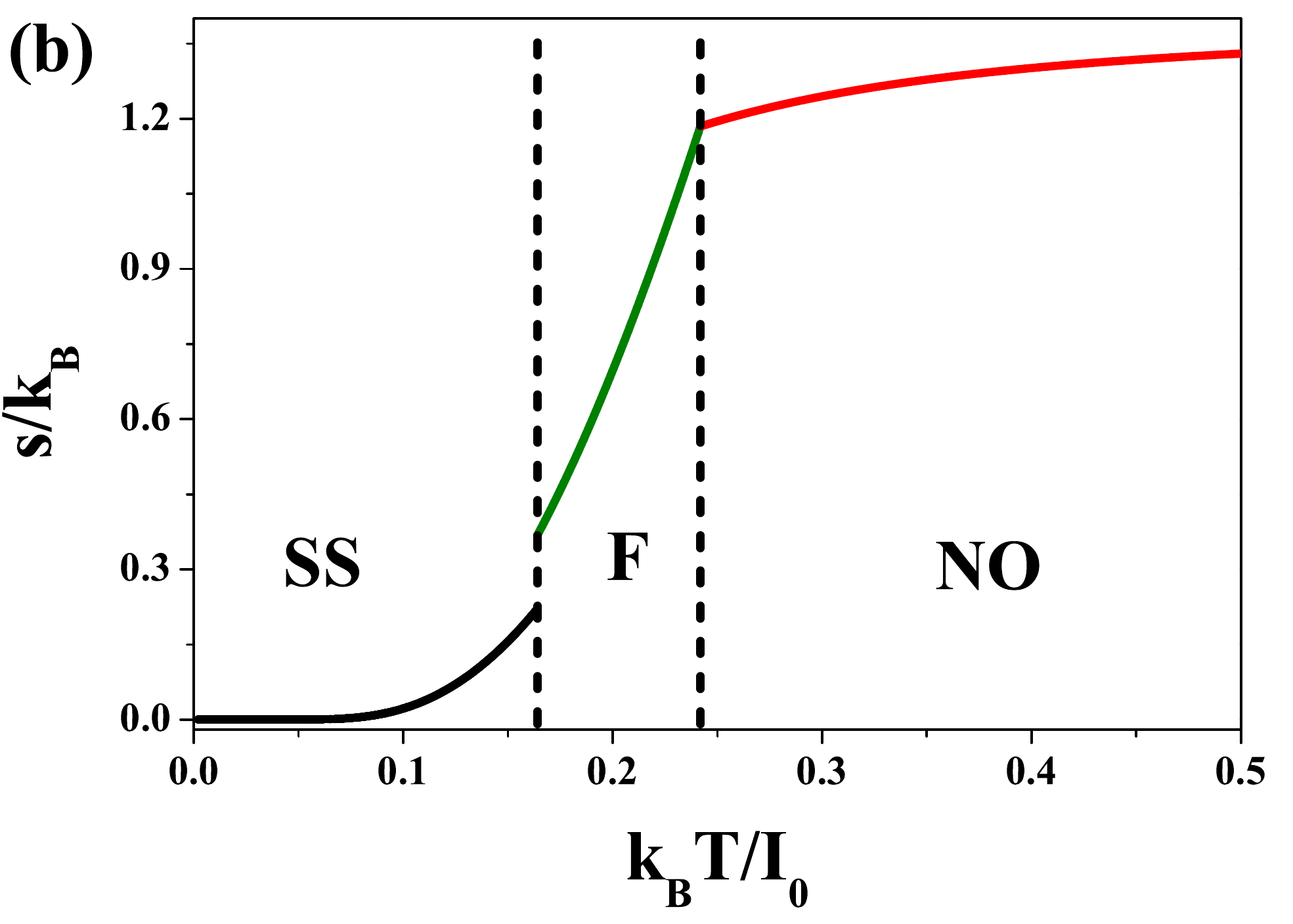}
    \includegraphics[width=\rozmiartrzy]{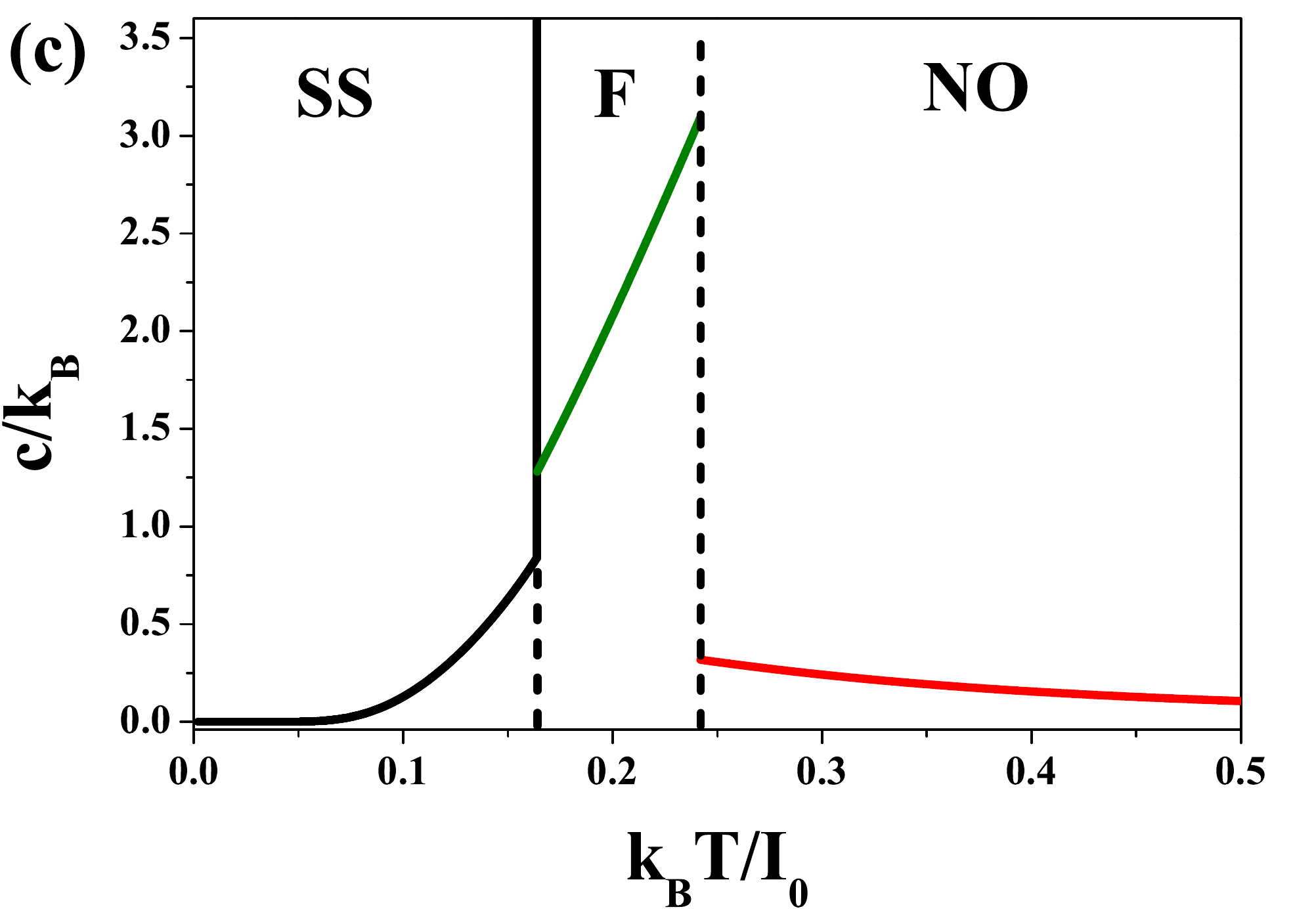}
    \caption{(Color online) Thermodynamic parameters (a)~the concentration of paired electrons \mbox{$n_p/n=2D/n$}, (b)~the entropy $s$ and (c)~the specific heat $c$ as a function of $k_BT/I_0$ for \mbox{$J/I=0.3$}, \mbox{$U/I_0 = 0.69$} and \mbox{$n=1$}.}
    \label{rys:parametry}
\end{figure}

The entropy $s$ and the specific heat $c$ can be derived as:
\begin{equation}
s  =  -\frac{\partial f}{\partial T}, \qquad c  =  -T\frac{\partial^2 f}{\partial T^2}.
\end{equation}

The temperature dependencies of the thermodynamic parameters
for \mbox{$J/I=0.3$}, \mbox{$U/I_0 = 0.69$} and \mbox{$n=1$} are shown in Fig.~\ref{rys:parametry}.

The concentration of paired electrons \mbox{$n_p=2D$} (normalized to the total electron concentration $n$) as a~function of temperature is presented in Fig.~\ref{rys:parametry}a.
At the \mbox{SS--F} transition large amount of electron pairs is destroyed. Thus $n_p$ has a sharp break at the \mbox{SS--F} transition temperature $T_{c1}$ and a~substantial fraction of single particles exists above $T_{c1}$. As temperature is lowered, the condensate growths  both from a~condensation  of pre-existing pairs and from binding and condensation of single particles.
At the \mbox{F--NO} transition (at $T_{c2}$) $n_p$ is continuous. In the NO phase it increases to \mbox{$n_p\rightarrow0.5$} at \mbox{$T\rightarrow+\infty$} (two electrons at the site is one of four equal probable configurations at the site and \mbox{$n=\langle\hat{n}_i\rangle=1$}).

The temperature dependencies of the entropy $s$ and the specific heat $c$ are shown in Figs.~\ref{rys:parametry}b and~\ref{rys:parametry}c, respectively.
$s$ increases monotonically with increasing $T$. At $T_{c1}$ the entropy $s$ is discontinuous whereas it is continuous at $T_{c2}$.
One can notice that in the high-temperature limit the entropy \mbox{$s/k_B\rightarrow \ln(4) \approx 1.386$} (there are four possible configurations at each site).
The peak in $c(T)$ is associated with the first order transition (at $T_{c1}$), while the \mbox{$\lambda$-point} behavior is typical for the second order transition (at $T_{c2}$).

\section{Final remarks}

We have studied a~simple model of a~magnetic superconductor with very short coherence length (i.~e.~with the pair size being of the order of the radius of an effective lattice site) and considered the situation where the single particle mobility is much smaller than the pair mobility and can be neglected.

One has found that the system considered for \mbox{$n=1$} exhibits various multicritical behaviors (determined by the ratio $J/I$) including tricritical, critical-end and bicritical points. It has been shown that, depending on the values of interaction parameters, three homogeneous phases: superconducting, \mbox{(anti-)}ferromagnetic and nonordered occur on the phase diagrams of  model (\ref{row:ham1}) at half-filling. The transitions between ordered phases (SS, F) and the NO phase can be first order as well as second order ones, whereas the \mbox{SS--F} transition is first order one. For \mbox{$n\neq1$} several types of phase separated states could be also stable in definite ranges of model parameters~\cite{KR0000}.

The other result of the interplay between magnetism and superconductivity could be appearance of triplet pairing~\cite{AGL2008}. Such a solution could appear together with ferromagnetic spin ordering. However, in model (\ref{row:ham1}) which assumes \mbox{$t_{ij}=0$} such a~state cannot be found. To investigate the possibility of occurrence of a~superconducting state with triplet pairing, the model should be extended to the case of finite bandwidth (\mbox{$t_{ij}\neq0$}) and be analyzed taking into account intersite pairing (in particular triplet pairing), e.~g. using Hartree-Fock broken symmetry framework~\cite{AGL2008,CR2006,MRRT1988}.

The mean-field approximation used to the intersite term is best justified if the $I_{ij}$ and $J_{ij}$ interactions are long-ranged or if the number of nearest neighbors is relatively large.
The derived VA results are exact in the limit of infinite dimensions \mbox{$d\rightarrow+\infty$}, where the MFA treatment of the intersite interactions $I$ and $J$ terms becomes the rigorous one.

Let us point out that in the MFA, which does not take into account collective excitations, one obtains the same results for the \mbox{$U$-$I$-$J^z$} model, i.~e. model (\ref{row:ham1}),  and the  \mbox{$U$-$I$-$J^{xy}$} model, where the term \mbox{$2J\sum{\hat{s}^z_{i}\hat{s}^z_{j}}$} is replaced with \mbox{$J\sum{(\hat{s}^{+}_{i}\hat{s}^-_{j}+\hat{s}^{+}_{j}\hat{s}^-_{i})}$}, describing interactions between $xy$-components of spins at neighboring sites, \mbox{$\hat{s}^{+}_i = \hat{c}^{+}_{i\uparrow}\hat{c}_{i\downarrow} = (\hat{s}^-_i)^\dag$}. In both cases the self-consistent equations have the same form, only a~magnetization along the $z$-axis becomes a~magnetization in the $xy$-plane~\cite{KKR2010}.

\begin{acknowledgments}
The author is indebted to Professor Stanis\l{}aw Robaszkiewicz for very fruitful discussions during this work and careful reading of the manuscript.
The work has been financed by National Science Center (NCN) as a research project in years 2011-2013, under grant No. DEC-2011/01/N/ST3/00413.
We would also like to thank the European Commission and Ministry of Science and Higher Education (Poland) for the partial financial support from European Social Fund -- Operational Programme ``Human Capital'' -- POKL.04.01.01-00-133/09-00 -- ``\textit{Proinnowacyjne kszta\l{}cenie, kompetentna kadra, absolwenci przysz\l{}o\'sci}''.
\end{acknowledgments}

\end{document}